Do Most Students Need In-Person Lectures?

A Study of a Large Statistics Class

Ellen S. Fireman[1], Zachary S. Donnini[2,3], Michael B. Weissman[4] , and Daniel J. Eck[1]

[1] Department of Statistics, University of Illinois at Urbana-Champaign

725 S Wright St., Champaign, IL 61820

[2] University High School, University of Illinois at Urbana-Champaign

1212 W Springfield Ave, Urbana, IL 61801

[3] Yale University

505 College St., New Haven CT 06520.

[4] Department of Physics, University of Illinois at Urbana-Champaign

1110 W Green St., Urbana, IL 61801

3/11/23

Abstract:

Over 1100 students over four semesters were given the option of taking an introductory undergraduate statistics class either by in-person attendance in lectures or by taking exactly the same class (same instructor, recorded lectures, homework, blind grading, website, etc.) without the in-person lectures. Roughly equal numbers of students chose each option. The online lectures were available to all. Attendance by online students was rare. The online students did slightly better on computer-graded exams. The causal effect of choosing only online lectures was estimated by adjusting for measured confounders, of which the incoming ACT math scores turned out to be most important, using four standard methods. The four nearly identical point estimates remained positive but were small and not statistically significant at the 95% confidence level. Sensitivity analysis indicated that unmeasured confounding was unlikely to be large but might plausibly reduce the point estimate to zero. No statistically significant differences were found in preliminary comparisons of effects on females/males, U.S./non-U.S. citizens, freshmen/non-freshman, and lower-scoring/higher-scoring math ACT groups.

## Introduction

Expanding the enrollment in large college classes through the use of online components has some obvious economic advantages. Dropping the requirement of in-person lectures could be a particularly important change because enrollment in large in-person lectures is limited by classroom availability and by the availability of enthusiastic expert lecturers. For students, in-person lectures often run into conflicts with schedules for other classes, for jobs, and for family obligations, making online lecture options desirable so long as they are not educationally inferior. The recent pandemic has led to other obvious online advantages and thus increased interest in using online lectures.

There is a common impression that online education is somewhat inferior to in-person education (Loeb, 2020) largely based on secondary education results, e.g. (Heppen et al., 2017). On the other hand, one meta-analyses(Allen et al., 2004) of college-level courses concluded that some online modes offered slight advantages, although with substantial variability between different methods lumped under the name "distance learning". Another meta-analysis (Means, Toyama, Murphy, Bakia, & Jones, 2010) that included data on K-12 classes found benefits of online components particularly if mixed with in-person components. Research on the factors favoring different versions of online methods is ongoing, e.g. (Joosten & Cusatis, 2019).

Although these prior studies explore a variety of online methods, in this work we explore only a well-defined narrow question in this field. The central question was what effect offering an online option had on student learning. Did the students who chose that option learn less or more than they would have otherwise? In the course we studied no new online material was added since most course materials (e.g. homework) were already online. Thus our work isolates the effects of a particular change- dropping in-person lecture enrollment- rather than looking at more complicated changes in multiple aspects of courses. As we shall show, the students who chose to enroll in the online option almost never attended in person, so that if these students needed in-person attendance to learn that would show up as a reduction of their exam scores. This paper describes an estimated average treatment effect (ATE) on objective exam scores of enrolling in the online version vs. the in-person version.

## Background on Methods

The practical question we are trying to answer to guide future decisions is "what effect would offering an online lecture option in a course with in-person lectures have on how much students learn?" Given that many students offered that option did choose it, the question becomes equivalent to the counterfactual question about observations of past results: "would the students who signed up for the online mode have learned more or less if they had signed up for the in-person mode?" Since as we shall show the online students rarely if ever attended class, our results directly bear on the more generally interesting question of whether class attendance is needed for most students. The answer is important in deciding what options to offer.

Such “what if” questions are by definition causal (M. A. Hernán & Robins, 2020; Pearl, Glymour, & Jewell, 2016), so we shall not shy away from causal language (Miguel A. Hernán, 2018). Since, however, we were unable to randomly assign students to different lecture modes, addressing the uncertainties in drawing causal conclusions from observational data will be the central theme of this paper. We try to teach students to be forthright about both causal claims and their uncertainties, and hope to set a good example here.

There are two main stages of causal analyses of observational data. Given a set of covariates one can construct causal models including those covariates, adjusting parameters to fit the data to isolate the causal effects of interest. If an outcome difference between treatment groups remains when conditioned on all the relevant confounders, that difference is attributed to the effect of the treatment. The most reliable way to do that is to randomly assign treatment, breaking any backdoor paths linking the treatment to other causes. When random assignment is not possible this stage can still be done with modern tools (M. A. Hernán & Robins, 2020; Pearl et al., 2016). It turns out not to be problematic for our study.

Nevertheless, sources of bias outside the model can arise from unmeasured covariates and selection effects. Dealing with those requires a sensitivity analysis involving plausibility arguments that rely on some domain-specific knowledge (Liu, Kuramoto, & Stuart, 2013; VanderWeele, Tchetgen, & Halloran, 2014). This sensitivity analysis is inherently somewhat subjective, but at least the subjective elements can be stated explicitly to help readers decide on their plausibility. Since the covariates to which we have access are far from fully predictive of exam scores much of this paper will be taken up with a sensitivity discussion.

In any causal question one needs to distinguish between the "what if" answers for how outcomes depend on a particular treatment and questions about various mediating steps on the path from treatment to outcome (M. A. Hernán & Robins, 2020; Pearl et al., 2016). Understanding the role of mediators can be important for understanding ways in which outcomes could change in different circumstances, but mediators are not relevant to the simple question of what the effects of the treatment were. In this study keeping track of the behavior of students along the path from course registration to exam grades can matter for understanding the results and extrapolating to other circumstances, but not for evaluating the effect of the treatment observed.

In describing past results and throughout this paper we will follow the convention of labelling as "non-significant" any effects for which the 95% frequentist confidence intervals include zero, to avoid the distraction of considering too many effects of unknown sign. From a Bayesian perspective, we do not, however, mean to imply a non-zero prior probability for any null hypothesis. In other words, some null hypotheses (e.g. that the effects of a change in lecture mode are *exactly* equal for different subgroups) have zero prior probability, even though they may turn out to be consistent with the data (Amrhein, Greenland, & McShane, 2019).

We shall show that the modeling with measured covariates together with plausible sensitivity analyses for unmeasured covariates indicate a weak, statistically insignificant positive effect of choosing the online option for those students who chose it. Since we have no covariates that are useful for finding to what extent the students systematically chose the option most suited for themselves, we will have no reliable way of estimating what the online effect would have been for the large minority of students who did not choose it.

## Background on Online vs. In-person Lectures

It is not obvious *a priori* whether to expect online or in-person lecture delivery to be more effective in an undergraduate course. There are some reasons one might expect asynchronous online lectures to be more effective, especially in any cumulative course requiring understanding each step before moving on to the next. These include that students can:

1. Adjust pacing by replaying the parts with which they have difficulty and pausing just long enough for in-lecture exercises.
2. Listen when they're in the mood and take breaks if they have trouble concentrating for 50 or 80 minutes.
3. Make up lectures missed due to emergencies.
4. Use closed-captions if they prefer them to spoken English.

Some disadvantages of asynchronous online lectures may include that students:

1. Lose the direct sense of personal involvement and interaction with other students.
2. Lose the chance to ask questions during lecture.
3. Face distractions in a non-classroom environment.
4. Are tempted to procrastinate, cramming multiple lectures shortly before exams (Figlio, Rush, & Yin, 2013; He, Gajski, Farkas, & Warschauer, 2015).

Whether the online advantages outweigh the disadvantages is likely to depend on course content, structure, student characteristics, and lecture style.

Simple comparisons of outcomes for online and in-person students do not give the causal effect of the teaching mode, since without random assignment the students in each mode may systematically differ. Adjustments for differences between the students enrolled in different modes is routinely done, e.g. in (Coates, Humphreys, Kane, & Vachris, 2004), and should be able to approximately correct for those differences if they are not too large and sufficient data on the students is available.

Several meta-analyses discuss comparisons of outcomes for different teaching modes, including in beginning statistics classes, but these include highly heterogeneous methods affecting many aspects of course structure and usually did not include either random assignment or careful adjustment for cohort differences (Allen et al., 2004; Mills & Raju, 2011). Nevertheless, there are a few well-controlled studies that compare outcomes for different lecture delivery modes, including a few randomized controlled trials (RCTs).

Most randomized studies have been conducted on economics classes. Arias et al. (Arias, Swinton, & Anderson, 2018) found that in-person material had a significant positive effect on some but not all outcome measures, when compared to specially prepared online material that did not include videos of the in-person lectures. (These authors also provide a good review of previous work.) Alpert et al. (Alpert, Couch, & Harmon, 2016) found that eliminating in-person material had negative effects on exam scores. They found that a blended online/in-person method had statistically non-significant negative effects compared to the in-person version (Alpert et al., 2016). Joyce et al. (Joyce, Crockett, Jaeger, Altindag, & O'Connell, 2014) found a modest (2.3 points on a 100 point scale) but statistically significant benefits of having two weekly in-person classes rather than one in a hybrid format. Figlio et al. (Figlio et al., 2013) found small benefits of using in-person rather than asynchronous online lectures. The overall effect became statistically significant when adjusted for random imbalance between the treatment groups. The benefits were concentrated in some subgroups, especially Hispanics (Figlio et al., 2013).

One particularly relevant randomized study found no significant effect on test scores in a beginning statistics class with the treatment difference consisting of reducing in-person lecture hours by about a factor of three (Bowen, Chingos, Lack, & Nygren, 2014). The authors note that although students were randomized, they were unable to randomize instructors, leaving a potential confounder (Bowen et al., 2014). The authors emphasized that the success of their “hybrid” treatment, with reduced in-person components, may have depended on the use of interactive online materials (Bowen et al., 2014).

Kofoed et al. (Kofoed, Gebhart, Gilmore, & Moschitto, 2021) have recently reported an RCT at West Point, showing that replacing 12-student in-person sections with 18-student synchronous online sections in a beginning economics course led to a statistically significant loss of about 0.22 standard deviations in overall grade. In the context of such small synchronous sections, we would expect that none of the potential benefits of asynchronous online lectures would occur, but that there would still be some loss of the classroom environment advantages, especially since the online sections were 50% bigger. Thus *a priori* one might guess that for large lectures available asynchronously online the balance of effects would shift somewhat toward favoring online benefits compared to the fairly small negative effect found in the West Point study.

Other fairly well-controlled studies include some early work on using recorded lectures in a statistics class that found little effect of switching from in-person lectures to in-class televised lectures, but with too small a sample to reliably find effects that might be practically important (Ellis & Mathis, 1985). Of the non-randomized studies with fairly stringent controls, one (Bettinger, Fox, Loeb, & Taylor, 2017) used instrumental variables (especially the intermittent availability of the in-person version of a class) to try to extract the causal effect of using a completely online version instead of a largely in-person version of a course taught at a large for-profit university. It found that the online effect on grades and subsequent performance in other courses was noticeably negative overall, especially for weaker students.  Since in-person classes were only taught when teachers specifically volunteered, while online classes were offered even when no teacher volunteered to be involved, there may have been systematic differences between the treatment and control teachers (Bettinger et al., 2017).

(Figlio et al., 2013) describe an economics course with similar structure and flexibility for students as our courses. They, however, temporarily reduced that flexibility for the students participating in their randomized trial in order to make a direct comparison of the two lecture modes rather than a comparison of the two flexible options. In our observational study the flexible options were maintained.

One study in an electrical engineering course looked at a flexible hybrid ("HyFlex") system (He et al., 2015), also used in some statistics courses (J. B. Miller & Baham, 2019; Jackie B. Miller, Risser, & Griffiths, 2013). Hyflex allows students to freely move between in-person and online modes, as did ours, but without each student registering at the start for one mode or the other. The finding was that students who more frequently chose not to attend in-person lectures did substantially worse on exams, after controlling for covariates representing some pre-existing student traits (He et al., 2015). It is difficult in such studies to tell how much the in-person attendance difference actually causes the outcome difference and how much both reflect motivational differences that the covariates do not capture.

The generalizability of the mixed conclusions of these varied studies to courses with different characteristics is unknown. Some of the key variables that they have suggested might moderate the online treatment effect are the nature of the treatment itself, which varies substantially between the different studies, and the strength of the students.

Educational Context

A brief description of the course studied may be useful for judging to which other courses our results might plausibly be generalized. Here we report results from four semesters of a large (~ 300 students/semester) beginning undergraduate statistics course offering an asynchronous online lecture option at a large selective public university. The course is a general education accelerated introduction. It has no calculus prerequisite but students are expected to be comfortable with algebra. Students come from a wide variety of majors, including psychology, economics, mathematics, various engineering and physical science fields.

The course relies heavily on the lectures since it has no discussion sections, no laboratories, and no standard text. Since there are no discussion sections many online elements were already included before the online lecture delivery option was started. Almost all elements of the course were shared by both online and in-person students. These shared elements include:

1) an incomplete-notes workbook/text filled out during lectures

2) frequent automatically graded randomized homework exercises (24 in a 15 week course) linked to an online discussion board and to relevant lecture passages
3) exams and practice exams, designed at least as much for teaching purposes as for evaluation
4) prelecture videos, recorded both by regular staff and by some outstanding undergraduates, with embedded bonus exercises
5) access to a user-friendly statistical computing program (http://www.istics.net/DataProgram/ )
6) in-person office hours, although these were not used much
7) anonymous surveys used to generate fresh data for analysis
8) general course website
9) email notifications.

For the first two semesters included in this study there was a nominal prerequisite of having some prior statistics course. The material was reorganized after the first batch of two-semesters both to improve the logical flow and to allow dropping the nominal prerequisite, since the variety of coverage in different prior statistics courses required reviewing essentially all the beginning material despite the prerequisite. In between these two two-semester batches, there was a semester in which the online and in-person versions were run by two different instructors, so we did not analyze data from that semester.

The cost of adding the online option was minimal, or perhaps negative, since the lectures were already being recorded for other purposes. The recordings were already used to help the lecturers evaluate and improve their own lectures, to provide new lecturers with guidance, and to serve as back-ups when students had to miss lectures.

Other than the scrupulous avoidance of differences (e.g. hints on exam content) between the treatment of the two groups, the course was run exactly as it would have been if no research had been anticipated. No changes were made for research purposes even to the anonymous online surveys.

## Treatment Methods

It was not possible to arrange random assignments of students to the two versions, but fortunately the two groups overlapped enough on the measured covariates to allow several methods of adjustment for the differences. All course materials made it clear that exams and homework would be identical for both groups, avoiding any impression that either version would be particularly easy. According to responses to an anonymous survey question asked in three of the four semesters for which we have exam data at least half the students said they chose the online mode for scheduling reasons rather than because they expected to learn better in that mode. Thus the online group contains many students who would have been likely to take the in-person version under slightly different circumstances.

Our intent at this stage was only to estimate the ATE over the whole collection of students. Our data were underpowered for a determination of different effects on most subgroups, e.g. ethnic categories. Nevertheless, due to increased interest in using online methods, we include a first look at interactions of the online treatment with several covariates- freshman status, math ACT scores, gender, and U.S. citizenship. These were the covariates about which we heard specific concerns from colleagues and for which there were reasonably large fractions of students in the relevant categories.

*The treatment under study consists of choosing not to enroll in an in-person lecture while keeping all the other course components unchanged.* In practice, as we shall see, this enrollment choice usually led to actual non-attendance of the lectures. The in-person group also had unrestricted access to the asynchronous lectures. In most previous work, e.g. (Arias et al., 2018; Figlio et al., 2013), access to online resources was at least nominally restricted for in-person students, although in some cases the researchers suspected that in-person students were obtaining access to the online material. Often, substantially different presentations were used for the online group (Coates et al., 2004). Coates et al. (2004) maintained flexible student options in their study, but the materials offered did not resemble the ones in our study. One previous study (Joyce et al., 2014) allowed the in-person group equal access to all the online material, but that online material included slides from but not videos of the in-person lectures.

The lectures were recorded at the in-person class and then posted online, usually within several hours, with occasional minor edits, e.g. removal of several-second portions in which a verbal slip had been made or in which the lecturer's phone rang. A typical verbal slip would be to incorrectly state the value of a number in an example, followed a few seconds later by a correction. In the first semester, however, recording difficulties occasionally requiring some office-recorded sections to replace missing in-class recordings.

Other than the lecture delivery, the one treatment difference between online and in-person versions was that a small amount of bonus credit was available to the in-person students who answered questions in lecture using an i-clicker, except for the first semester of the new version, Spring 2019,  when new i-clicker exercises were not yet ready. Although both groups received bonus points for completing their lecture note workbooks, more were given to the online students to approximately balance the i-clicker points. Thus those students who registered for in-person lectures had some extra motivation to frequently attend. The algorithm for counting bonus points was to add the bonus points to both the numerator and denominator of the student's numerical grade average, so that it has no effect on grades that were already 100%. Since it raises final grades by an amount approximately proportional to how far the non-bonus grade was from 100%, the bonus points primarily served to motivate students who were not otherwise doing well in the course.

The availability of the online lectures to the in-person students was a major part of the treatment under study. Thus use of the online material by the in-person students does not represent a crossover from one treatment arm to the other. Although use of the online lectures by the in-person group is therefore irrelevant to estimating the ATE, it is relevant for speculation on potential mechanisms for any treatment effects that might be found. As expected, almost all the in-person students reported making use of the online lectures, but on average only about 60% as frequently as the online students.

Attendance of in-person lectures by online students is also not a crossover between intended treatments, since ESF encouraged maximum student flexibility. Thus it is also irrelevant to estimating the ATE for the treatment used. It is, however, relevant to the *generalizability* of our results to other more typical treatments. If the online students ended up going to lectures about as much as the in-person students, then our results would only be relevant to an unimportant registration formality. As we shall show, in-person attendance showed a dramatic contrast between the groups, so that our results are relevant to evaluating the effects of in-person attendance, not just formal registration.

The anonymous survey results can be used to check the informal impression that online students hardly ever came to class. Bonus points offered for survey participation maintained high response rates, usually over 70%. We are not so naïve as to believe that even fully anonymous student self-reports of lecture attendance or viewing should be taken literally but they are adequate for qualitative comparisons.

About 80% of the online students reported never attending an in-person class. (We do believe this particular result.)  About 15%  reported attending once or twice. About 5% said they attended frequently. We think these 5% were mainly students who would have taken the in-person version except for scheduling glitches, such as part-semester courses that overlapped the lecture times. Thus we think that our online results should be nearly the same as would be found in a different treatment in which in-person lectures are not available to an online group, at least if it consisted of students similar to our online group. About  75% of the in-person students reported attending more than half of the lectures. Although that is probably an overestimate of the sharp contrast with the online group is  real.

Data collection stopped in Spring 2020, when only about 13% of the students in the main options chose the in-person version. (Other students, however, chose small in-person sections with lectures taught by teaching assistants. Many of those students said that they made use of the online lectures.) After Covid-19 hit mid-semester all students were transferred to the online version and we had to switch to an un-proctored open-book exam format. This forced switch would have provided an opportunity to estimate unmeasured confounding effects as well as effects on students who had not chosen the online version, if it had happened at the start of the semester, with reasonable numbers of students enrolled in both versions, and with reliable exams.  Unfortunately, none of those conditions held.

**Evaluation Methods**

The only evaluation we looked at for this project was performance on computer-graded exams, to avoid any subconscious bias on the part of graders. These exams also are our best measure of learning outcomes. Almost all the students took the exams in-person in mixed groups of in-person and online registrants. A small number of the online students (roughly 3%) and occasional in-person students took the same exams remotely, proctored via a commercial service using webcam monitoring.

The evaluation method changed between the two two-semester batches. The first batch used hand-graded midterm exams, which we did not look at for the purposes of this study, and cumulative computer-graded finals, the results used here. The second batch used three equal-weighted computer graded exams, with the semester average used here. We denote these objective exam scores ObjEx. We did separate analyses of these two batches. Despite the different forms of exam scores used, the standard deviations for all four semesters were similar, ranging from 8.8 to 10.5. (All scores presented here are on a 100 point scale, with a difference of 10 representing one grade point.) There was more variation of ObjEx means and standard deviations within two-semester batches than between them. To reduce statistical uncertainty, we combined all the data for an overall comparison of whether switching to all-online lectures affected objective exam scores in these two closely related batches of the same course.

In order to avoid unnecessary bias due to confounders, we used four methods to adjust for differences between the students in the two versions in order to estimate the causal ATE of dropping in-person lecture enrollment. The first three were all standard methods: multiple linear regression (MLR), stabilized inverse propensity weight (IPW) (Austin & Stuart, 2015; Lunceford & Davidian, 2004), and doubly-robust (DR) (Robins, Rotnitzky, & Zhao, 1994). We checked these for the overall results via a much more computation-intensive less-parametric outcome-highly-adaptive-lasso (OHAL) method (Ju, Benkeser, & van der Laan, 2020). We made no original contribution to these methods but hope that our diligence in using them will encourage other education researchers to do likewise.

We included as covariates those relevant predictors to which we had access and whose values were set before the treatment started. The campus data support service supplied anonymized data files with the relevant covariates. These included the following categorical variables: student year in school (four categories), semester when the course was taken (also four categories), gender (treated as dichotomous male or female in university records), and dichotomous U.S. vs. non-U.S. citizenship. The quantitative variables used were ACTmath and ACTverbal scores (including SAT equivalents), overall ACT score, high school GPA (HSGPA) and the approximate median ACT score of the major in which the student was enrolled. That ACTmajor was obtained by averaging the scores of the 25th and 75th percentiles, available from a university website for prospective students. The ACTmajor score was initially used as a proxy for ACT scores before those were available to us, but to our surprise remained a significant predictor even after individual ACT scores were included.

HSGPA values were obtainable for 61% of the students. The remainder of the students were from systematically different groups, especially transfer students and international students, who it was also important to include in the ATE estimate. Therefore we ran a full analysis omitting the HSGPA on the larger sample, which also had the advantage of reducing statistical error. To estimate any correction for systematic differences in the traits measured by HSGPA, we tested how much inclusion of HSGPA changed the ATE on the subsample for which HSGPA was available.

The initial exploration was via MLR using least-squares fitting, using both R and the same point-and-click program used in the class. The MLR method gives unbiased estimates if the linear effects model is correctly specified. The MLR residuals were non-Gaussian, and not constant across the predictor space, which is unsurprising given the highly left-skewed exam scores conditional distribution, so we checked confidence intervals using standard bootstrap methods (Faraway, 2014).

We used three additional methods to check that the adjustments are not too sensitive to any errors in that MLR model. The second adjustment method was a standard stabilized IPW analysis (Austin & Stuart, 2015; Lunceford & Davidian, 2004). IPW adjusts weights for different points to create a pseudo-sample in which the distributions of the covariates are approximately balanced between the groups. This method gives unbiased estimates of the ATE if the model for the propensity of different types of students to take each version is correctly specified. Our propensity score model used main-effect logistic regression on the same covariates as the MLR model. We checked that the important predictive covariates were fairly well-balanced in the pseudo-sample generated by the IPW method (Austin & Stuart, 2015).

The third method, DR, corrects for any imbalances in the IPW pseudo-sample by using MLR estimates to give unbiased estimates if either the linear effect model or the logistic propensity model is correct (Robins et al., 1994). The DR estimate of what score a student would have obtained if they had received the other treatment uses an MLR trained on students receiving that other treatment. DR is thus particularly suited to estimating counterfactual outcomes.

The fourth method, computation-intensive OHAL, can reduce problems from model mis-specification by trying a wide variety of different models, checking each by internal cross-validation (Ju et al., 2020). Since OHAL runs much more slowly than the other methods and since its advantages require large data sets, it was used only to check the key bottom-line ATE estimates. A close match between OHAL results and results of the simpler methods suggests that more complicated general models are not needed.

The ATE confidence intervals for the first three methods were estimated using standard bootstrap methods (DiCiccio & Efron, 1996). Confidence intervals on the small adjustment for inclusion of HSGPA were determined by a paired bootstrap analysis, in which the change of the ATE estimate from including HSGPA in the model was determined for each bootstrap sample. That allows the resulting small HSGPA adjustment to be made with little increase in the confidence interval width for the ATE. Confidence intervals for the OHAL procedure were obtained using cross-validated standard errors (Ju et al., 2020).   Full descriptions including the R code and full results for these methods are presented in the Supplementary Appendix. All calculations not shown in the Appendix were done without coding just using multiple linear regression in the point-and-click program.

1338 students (545 registered in-person and 793 registered online) were participating in the course ten days after the starts of the semesters according to university records. 1196 of them completed the course and took the final exam. We included all those in an initial calculation of the raw point difference between online and in-person. In order to calculate an ATE we needed a sample with suitable covariates to use for adjustments, so we dropped 91 students for whom no admission test scores were available or who were not enrolled for an undergraduate degree. This left 1105 students for the analysis sample, including 675 for whom we had HSGPA. This analysis sample included 91% of the in-person students for whom we had final grades and 93% of the online students with final grades. The overall raw score average difference between online and in-person groups was similar in the full sample (1.26) and the analysis sample (1.38).

The possibility of important unmeasured or inadequately measured confounders is typically the biggest potential issue for any non-randomized study attempting to estimate causal effects. We used the substantial variation in the covariate imbalance between semesters  to estimate possible under-correction for the traits measured by our covariates, finding no major effects. We were concerned about two potential confounders that our covariates would not pick up-  prior commitment to this particular course, as opposed to more general conscientiousness, and prior preparation for this particular course as opposed to general mathematical preparation.

With regard to prior preparation for statistics, in the first two semesters some unspecified prior statistics exposure was nominally required. In the second two semesters the routine anonymous online surveys showed very small differences between the fraction of online students (80%) and in-person students (77%) self-reporting having taken prior courses. Even neglecting the likely collinearity between this preparation covariate and others used, no plausible effect magnitude of this covariate could have an important effect on our estimated ATE, so we did not attempt to obtain transcripts to sort out such effects.

We suspected that students who chose to allot a scarce schedule slot to this course would be more likely to put effort into it than would students who avoided any constraints on their other courses by choosing the flexible online option. If so, our estimates of ATE would tend to favor the in-person version compared to the true ATE. Joyce et al. (Joyce et al., 2014)  found that causal effects of randomized assignments to lectures were much smaller than the covariate-adjusted association of lecture attendance students with outcomes among those not enrolled in a randomized trial. Thus the association of freely chosen lecture attendance with exam scores is mostly due to confounding by some motivational factors (Joyce et al., 2014). Confounding effects of this sort might lead to overestimates of the advantages of in-person lectures in non-randomized studies (He et al., 2015; Joyce et al., 2014). While we suspected that the effect would have the same sign in our observations, skipping lectures for which one has signed up is enough different from choosing not to sign up for them that we cannot use the prior result (Joyce et al., 2014) to make a quantitative estimate of the effect or even to be sure of the sign.  Instead, we used a comparison of our main outcome, exam scores, with a more effort-weighted outcome, homework scores, to try to estimate whether any effect of different commitments in the two treatment groups was likely to be important.

**Results**

According to university records 695 (87.6%) of the online students and 501 (91.9%) of the in-person students who were registered in the course on the tenth day of the semester completed it. These figures somewhat overstate the dropout rate because some of the students who were officially registered at that point either never did a class assignment or had already ceased to do any. The difference in dropout rates is significantly bigger than would be expected for a hypothetical equal systematic dropout probability (p=0.01). Table 1 shows the completion and dropout numbers for each mode in each semester. Both the rates for the first three semesters are fairly typical for math courses (Paden, 2006).

The larger dropout rates in the first two semesters were in part due to a deliberate policy of urging students with less than the expected prerequisite knowledge to switch to a less-accelerated course that had no statistics prerequisite. By F19 there was no prerequisite and the expectations for students had become more widely known. We think that is why the dropout rates were lower.

Table 2 compares the values of the covariates in the two treatment groups for the 1105 students for whom they were available. The US/international and male/female distributions were within the range that could have plausibly occurred even by random assignment. For the ACT scores and HSGPA the differences between the group means ranged from 8% to 18% of the SD for the individuals. The enrollment by college year differed substantially. Students said that freshman advisors strongly discouraged them from taking the online version, so the large difference in freshman numbers is not necessarily indicative of some more general treatment-choice-linked imbalance in unmeasured student characteristics. Juniors and seniors tended to take the online version more than sophomores did, perhaps for reasons similar to those that accounted for the ACTmajor imbalance. Overall, the differences between groups are clearly large enough that adjustments for these differences can be important in accurately estimating the ATE and that care must be taken to see that such adjustments are not highly model dependent.

The mean raw exam scores and their SDs for each semester are given in Table 3, along with the corresponding ATE estimates and their bootstrap confidence intervals. The score scale is 0-100, with 10 points corresponding to one grade point. The average raw score difference, correcting only for the difference in exam difficulty between semesters, is 2.06.

The bootstrap confidence intervals shown for the ATEs are calculated on the implicit assumption that the random error terms for individual students are independent. This should be a good approximation since almost all student interactions occurred outside lecture in the online discussion space shared by all students and even in-lecture interactions are viewed by all students., i.e. there was no venue for cluster effects.

Between-semester effect variation can arise from between-semester variations in the treatment and in imbalances in student traits if the model fails to fully adjust for those. We can compare the variation found with that expected under the null model of a uniform true effect for each semester with uncorrelated individual random effects. We compare the variation in the $ATE_{MLR}$ for the four semesters with that null using the good approximation that the individual random variations give normally distributed ATE estimates with widths estimated from the confidence intervals. The result is a 3-dof chi-squared of 6.8, with p=0.08, not sufficient to confidently reject the null. $ATE_{DR}$ and $ATE_{IPW}$ give similar results.

Although the excess chi-squared over the null expectation is not quite statistically significant at the conventional alpha=0.05 level, it is large enough to raise the question of whether some explanations can be found. Of most concern, the mix of students in the two versions changed over time, as can be seen in the changing adjustments for covariates in Table 3. Any failure of the model to adequately adjust for differences between groups would show up as between-semester variation in ATE anti-correlated with the adjustments, just as we see. The main problem with any such failure would not be mis-estimation of the statistical uncertainty but the first order effect of mis-estimation of the ATE itself, an issue to which we will return in the sensitivity analysis.

There can also be effects of slightly different treatments in the different semesters. The most positive ATE was found for Sp19 for which there was no i-clicker bonus incentive to attend lectures. The only negative ATE was found for Fa17, the first semester in which the recordings were used in this course, when there were some problems with getting good-quality recordings out quickly and three had to be re-recorded in an office. In the first two semesters students were not sent emails with a reminder of the weekly schedule. Such minor changes in the treatment are not part of the statistical uncertainty in estimating the average effect of the treatments we used but rather minor indicators of something that would be assumed anyway- that extrapolating any educational results to other classes is uncertain because class structures and contents differ.

The ATE could differ between the versions of the course used in the two-semester batches. In the single semester data, however, most of the variation was found within those batches rather than between them, so in order to get good statistics we lumped all four together, under the plausible assumption that the net ATE varied little between semesters. Table 4 gives the multiple regression, stabilized IPW, doubly-robust. And OHAL estimates for the ATE for the combination of the four semesters, along with the 95% confidence intervals. The four overall ATE estimates are nearly identical, and with almost the same confidence intervals. Thus there is no indication that the more general models explored by OHAL are needed.

Inclusion of HSGPA in the model on the subsample of 675 students for which it was available increased $R^2$ from 0.32 to 0.37 on that subsample. It increased the ATE in this subsample by 0.11, 0.09 and 0.15 for MLR, IPW and DR methods, respectively, each with 95% confidence intervals of ~±0.4. Assuming that the traits measured by HSGPA have roughly similar effects and similar group differences in the 39% of the sample for which HSGPA was not available, the corrected ATE estimates would then be 0.74, 0.83, and 0.73 for MLR, IPW and DR methods, respectively. Assuming the errors add in quadrature each 95% CI would be just under ±1.2.

The overall point estimate for the online effect is positive, but with each estimate close to 0.75 points out of 100 it is small for practical purposes and not statistically significant at the 95% confidence level in this sample.

We explored interaction effects of the treatment with the variables suspected of being relevant to the effectiveness of the online treatment (gender, citizenship, freshman status and ACTmath). Adding these effects, either individually or all simultaneously, gave no significant interaction term. We also tried stratification on the same variables, converting ACT math to upper half and lower half categorical strata. Stratification both gives a slightly different look at possibly inhomogeneous effects and a way of checking robustness of the model. The online $ATE_{MLR}$ remained non-significantly positive for both US and non-US citizens, both freshmen and non-freshman, and both the upper and lower halves of the ACTmath distribution. The difference between the stratified results for the genders was also not significant. The stratified estimated $ATE_{MLR}$ was non-significantly negative for females and nominally significantly positive (p=0.035) for males. Due to the multiple comparisons the effect for males should not be considered to be shown to be positive with confidence, particularly since the opposite sign of dependence on gender was found in a related study (Figlio et al., 2013). We looked at the stratum of upperclassmen to check for model robustness rather than out of concern about that group. We again found a non-significant positive effect.

Although the ACTmath SD was 4.1, enough range to easily see its predictive power, less than 1% of the students scored below 20, the approximate national median. Thus although we saw negligible dependence of the effect on ACTmath we have essentially no evidence on how well the online treatment would work in a course like this for students with ACTmath scores below 20.

Since all methods gave nearly the same point estimates and confidence intervals we may use the MLR results, whose coefficients have simple familiar intuitive interpretations, in most of the remaining discussion. Table 5 shows the multiple regression coefficients for the covariates of the MLR model on the full analysis set of 1105 students, for which $R^2$=0.34. ACTmath is the most significant covariate.

Sensitivity Analysis

Despite the excellent agreement of the different mathematical methods of estimating ATE from the scores and the covariates there are reasons to worry about unmeasured or inadequately measured confounders. We have used covariates that give good indicators of students' general strengths but none that are particularly sensitive to how much effort they would be likely to put into this particular course. Furthermore, the correlation between the difference in ATEs between semesters and the differences in the adjustments to the raw scores looks suspicious, since it might indicate that the adjustments for imbalanced covariates are underestimated (Draper & Smith, 1998). Since the overall adjustment reduces the online benefit, further adjustment along the same lines would reduce it further.

We used several methods to estimate the plausible range of values of corrections for potential unmeasured confounders (UC). These were:

1. Simple comparison of the Cohen d of the estimated ATE with the Cohen d's of the quantitative variables, after correction for categorical variables.
2. Estimate of the correction for unmeasured quantitative confounders if they are as unbalanced as the measured ones and accounted for as much of the missing $R^2$ as plausible.

3. Use of the substantially different imbalances in pre-existing student traits in different semesters, including changes in sign, as an instrumental variable (Pokropek, 2016) to set plausible limits on further corrections for imbalances.
4. Use of the motivation-sensitive outcome variable, HW, to see effects of course-specific motivation, although at the risk of contaminating the causal estimate by inclusion of a variable downstream from the treatment. Including HW as a covariate explained most of the between-semester variance.

The full arguments are somewhat complex and lengthy, so we have been put them in a Supplementary Appendix. We may summarize the results here. The informally plausible range of the point estimates for ATE adjusted for confounding by inadequate measurement of general traits ranged from 0.08 to 0.85. The statistical error bars of ±1.1 remain. Plausible adjustment for course-specific motivation might further lower the estimates by as much as 0.3.

The key assumption shared by these estimates is that the unmeasured confounders are about as unbalanced as the measured ones. A qualitative glimpse can help to see if that is reasonable. We have survey data for three semesters on whether students "like math". On average, the in-person students said they liked math more, with the difference being strongest in Fa17. This attitudinal variable then could help account for the different ATE found in Fa17. That weakly suggests that missing motivational confounders might bias the results slightly to favor in-person, as we initially suspected based on prior work. In general, however, noise in measurements of covariates leads to under-adjustment, which in this case would slightly bias the results to favor online (Draper & Smith, 1998).

One cannot rigorously exclude the possibility of some important missing confounder, but if one existed it would have to meet some stringent requirements. Readers can judge for themselves the plausibility of finding important covariates that meet three conditions:

1. Exist at registration time.
2. Are orthogonal to the variety of covariates already in the model, including HW.
3. Are substantially more imbalanced than the covariates we have.

The different dropout rates between the online and in-person groups can also create a bias in estimating ATE. If the extra 4.7% dropping out from OL dropped out were concentrated near the pass-fail border with exam scores ~25 points lower than the mean, then that would shift the raw score advantage for the online group up by 3.9% of 25, i.e. 1.2 points. We think that is a seriously high estimate because many students who drop out have personal reasons unrelated to their expected grade, e.g. illness or having taken on an overly ambitious schedule. If we had a perfect model, with R=1, the resulting ATE would be unaffected by dropouts since the predictive coefficients on the remainder would be unchanged. We have an imperfect model with $R^2$=~0.4, so in this extreme case the ATE would be biased by ~0.7 points in favor of the online treatment. If, however, a more predictive model using better-measured covariates were to lower ATE that would reduce any such selection bias.  The plausible bias due to dropouts is nearly the same as the plausible bias due to missing confounders, which we estimated above. Since inclusion of missing confounders would remove potential dropout selection bias these two potential biases cannot be combined. We note also that most of the gap in dropout rate occurred in F17, so that the semesters in which the estimated ATE was positive had smaller potential selection bias, under 0.27 points.

To summarize the sensitivity analysis, barring some unlikely coincidence there will be unmeasured confounding affecting our ATE. Given that including the available covariates slightly reduced ATE, it is reasonable to guess that a more complete set would reduce it a little more. Plausible estimates of such effects are constrained to be small. Since the Fall ATEs (upwardly adjusted for quantitative covariates) and the Spring ATEs (downwardly adjusted for quantitative covariates) all lie within a narrow interval, the additional systematic uncertainty should be small even if the adjustment magnitudes are underestimated. One fragment of evidence on a suspected confounder ("like math") even points toward increasing the true ATE.  Dropout selection bias could conceivably be about as big as unmeasured confounding, but eliminating unmeasured confounding would almost eliminate selection bias, so these two potential corrections cannot both be added.

## **Discussion**

We may ask how much of the increased drop-out rate in the online group (12.4%) vs. the in-person group (8.1%) was caused by the different lecture modes. One would expect that students who had set aside a schedule time for the class might already be less prone to dropping out than those who had not made that commitment. The drop-out rate from the peak initial enrollment of 368 in Spring 2022 (the last semester taught by ESF) was only 3.8 % even though all students were enrolled online. Thus we do not think that the online mode itself causes much if any increase in drop-outs, since if it did the overall dropout rate would be higher when all students were online. The online mode was just a better option for students who wished to tentatively try the course without committing a schedule slot to it.

The treatment under study consisted simply of letting students choose not to enroll in in-person classes while keeping other course components fixed, and most online students did in fact never attend classes. Thus it might seem surprising that a negative effect was not found, since the general presumption is that removing resources will usually have at least a small negative effect, as others have found for switching to all-online lectures (Alpert et al., 2016; Arias et al., 2018; Coates et al., 2004; Figlio et al., 2013; Joyce et al., 2014). One might expect particularly strong lecture-mode dependent effects in a course without laboratory or discussion sections and without a complete text. There are several potential explanations for our results.

Although the point estimate of the effect was weakly positive the statistical and systematic CIs are large enough that we cannot rule out the possibility that dropping in-person enrollment had a small negative effect. Even allowing for plausible systematic uncertainties the actual effect is quite unlikely to be negative enough to have practical significance.

There is also a possible mechanism by which signing up for the online version could have caused a slight positive effect, consistent with our point estimate. The students who signed up for in-person lectures accessed the same lectures online about 40% less often that those who signed up for the online version. This difference in online viewing was likely to have a large causal component, i.e. one would expect that attending an in-person lecture would reduce a student's motivation to spend time on the recorded version. Since, for reasons stated in our introduction, asynchronous online lectures may provide a more efficient use of student time than do in-person lectures, it is possible that the true treatment effect of dropping enrollment in the in-person lectures could have been weakly positive.

The absence of significant interaction effects with freshman status or with ACTmath scores may seem more surprising than the statistically non-significant net effect estimate, since previous work has generally indicated that better-prepared or stronger students have better online effects (Bettinger et al., 2017). Most of the students taking this class were already comfortable with some mathematics. For the larger Course A, for which informal raw results were qualitatively similar, students had a much broader range of math preparation. Nevertheless, almost all students even in that course had been selected into a highly-ranked public university, so we do not know how well these methods would work for students who could not get in to such a university. Given the restricted range of our sample, the lack of an interaction term with ACTmath is entirely compatible with previous results (Bettinger et al., 2017). Future work might examine whether the hint of different effects for males and females is reproducible.

Although our results were underpowered for looking at interaction effects of the treatment even with measured covariates, presumably there are some differences among students in the relative value of online and in-person lectures. Students may have tended to take the version most suited to produce good test results for themselves. To the extent that such effects may be present, the estimated ATE would not give an accurate prediction of the net effect of requiring *all* students to take one version or the other. Our results show that it is unlikely that *allowing* most students to choose the online version of the lectures led to any important loss of learning,. We cannot rule out that requiring the minority who chose the in-person version to take the online version might have reduced their learning. An RCT finding ATE for randomly assigned students at some school where registration procedures allow it could find how big any such effect is.

Since the online version still offered in-person office hours, it was not exactly equivalent to a fully remote online version. These office hours were not used much, however. Most students never used them, and only ~10% of the students reported using them more than “once or twice” on the surveys. Summer sessions of the course, taken remotely, function well without them. After Covid hit they were replaced with Zoom versions. We think that this change is not very important.

As with any educational method, the generalizability to other circumstances cannot be rigorously determined, since it depends not only on details of the treatment but also characteristics of the course, the students, and the instructors. Nevertheless, some discussion may help readers estimate the relevance of our results to their courses. Results might differ in either direction for lecturers other than ESF. One aspect of this course- the absence of any discussion sections- may have made it more difficult for the online version to succeed than in courses for which sections would provide in-person contact regardless of lecture mode. On the other hand, there are several aspects of this course that may tend to make the online version work particularly well and thus may account for why our results are a bit different from the previous studies (Alpert et al., 2016; Arias et al., 2018; Figlio et al., 2013; Joyce et al., 2014) that found advantages for in-person lectures.

1. This is a cumulative logical/mathematical course, which may make some asynchronous online benefits especially relevant, especially the ability to go back over difficult steps.
2. The incentive to fill out the incomplete notebook may have reduced the effect of environmental distractions by keeping students actively engaged in lecture.
3. Nearly all of the students in this course may have been well enough prepared and web-savvy to avoid some detrimental effects of missing in-person lectures.
4. Perhaps most importantly, having a lecture-linked homework assignment due shortly after every lecture except pre-exam reviews avoids the biggest problem that has been cited (Figlio et al., 2013; He et al., 2015; J. B. Miller & Baham, 2019) for asynchronous online versions- the tendency of students to delay watching the lectures until cramming before exams.

Although we suspect that for many other STEM courses asynchronous online lecture delivery will be as effective for most students as in-person lecture delivery, we reiterate that we are not claiming that in-person course components add little value to online components as a general rule. Since most large courses have discussion sections as well as lectures, it would be useful to investigate whether offering an online section option would work about as well as in-person ones for those students who choose it, allowing students to participate remotely. Comparisons could use techniques like those we used here so long as sufficiently complete covariates are available. Whether replacing all in-person sections with online ones would work well could best be studied by randomized assignments, as in (Kofoed, Gebhart, Gilmore, & Moschitto, 2021), but with equal numbers of students in each type of section.

We have made no effort to compare with a conventional purely in-person version lacking online lecture backup. Since the in-person students also made use of the asynchronous online lectures, we suspect that a version without them would be inferior to either of the versions we examined. . From conversations with colleagues we know that many instructors prefer to give in-person students online access only for lectures missed for reasons such as illness. That limitation may be helpful if needed to keep enough lecture attendance to maintain a good classroom atmosphere. In-person lecture attendance might also be increased by increasing the credit for i-clickers or other attendance measures. Lecture attendance by online students was rare enough that we do not think that its lack has any notable consequences for purely remote students.

Since students tend to adjust their effort to meet the goals they set for their grades, exam grades can mask other important outcomes such as whether the class was a pleasant and efficient use of the students' time.  Other than the enrollment figures, we have only a little quantitative information on how students felt about the online option. In anonymous surveys at the ends of the Fa17, Sp18, and Fa19 semesters students were asked if they wished they had taken the other mode. Of the in-person students 21%, 44%, and 49% said yes in those semesters. The corresponding numbers for the online students were 25%, 26%, and 21%.  Thus overall students seem to be happier with the online mode than with the in-person mode, although the response is hardly unanimous. One of us (ZSD) took the class online as a high school student due to schedule constraints and was among the students who was grateful for that opportunity.

Although  we found no evidence that the students needed the in-person experience, we did not test whether the lecturer needed it. The course was developed over several semesters with live feedback as the lectures were delivered to in-person classes. The lecturer here (ESF) subjectively feels that live feedback was important and perhaps essential for the course development. Anecdotally, many other lecturers tell us that they need the in-person experience to lecture well. ESF's subjective experience in the Covid era also supports the idea that regardless of how the students view the lectures, it is harder to lecture well without immediate live feedback.

Although we strongly suspect that the indirect effect of lecture mode on learning via its effect on the instructors may be more important than the direct effects on the students we have no hard data on that issue. A future project might study the indirect effects via the effects on the lecturer by comparing outcomes for students receiving online lectures that were recorded in class to those for students receiving lectures that were recorded without a live class.

As an epilog, we add that this course remains all online for now due to health constraints. In Spring 2022 354 of the 368 originally enrolled students completed the course. A larger (~1400 students/semester) less accelerated course with similar structure has resumed offering an in-person option, which attracted fewer students than the online version even before its seats were filled. Many students continue to vote with their feet for the online version.

The R-code, more detailed results, and detailed sensitivity analysis are available as a supplementary file on request to the corresponding author.

**Acknowledgments:** We thank Yuk-Tung Liu for essential technical assistance, Kurt Tuohy from the UIUC Atlas team for crucial timely help in obtaining anonymized covariates, Jing Wang for helping analyze initial data, and Karle Flanagan for informative conversations about similar versions of another course. The UIUC Institutional Review Board authorized a waiver for this project since it is a purely retrospective study of teaching success not involving any added activities on the part of the students and uses only fully anonymized records.

**Conflict of Interest Statement**: One of us (ESF) is an author of the incomplete notes used in both versions of the course under study, and could receive royalties in the unlikely event that either version were adopted elsewhere. One of us (ZSD) personally benefitted from taking the online version as a high school student.

| Semester | OL completion | IP completion | Difference |
|---|---|---|---|
| F17 | 147/175=84.0% | 146/157=93.0% | 9.0% |
| Sp18 | 169/194=87.1% | 128/146=87.7% | 0.6% |
| Sp19 | 216/250=86.4% | 79/89=88.8% | 2.4% |
| F19 | 163/174=93.7% | 148/153=96.7% | 3.0% |

**Table 1.** Completion rates.

| Trait | N | $N_{OL}$ | $N_{IP}$ | Mean | SD | $Mean_{OL}$ - $Mean_{IP}$ | Cohen d |
|---|---|---|---|---|---|---|---|
| ACT | 1105 | 647 | 458 | 30.5 | 3.5 | 0.38 | 0.11 |
| ACTmath | 1105 | 647 | 458 | 32.4 | 4.0 | 0.45 | 0.11 |
| ACTverbal | 1105 | 647 | 458 | 58.8 | 8.0 | 0.60 | 0.08 |
| ACTmajor | 1105 | 647 | 458 | 30.2 | 2.6 | 0.48 | 0.18 |
| HSGPA | 675 | 377 | 298 | 3.52 | 0.35 | 0.036 | 0.10 |
| | | | | | | | |
| | | | | | | % of OL | % of IP |
| International | 435 | 270 | 165 | NA | NA | 42% | 36% |
| Female | 474 | 290 | 185 | NA | NA | 45% | 40% |
| Freshman | 155 | 40 | 115 | NA | NA | 6% | 25% |
| Sophomore | 407 | 212 | 195 | NA | NA | 33% | 43% |
| Junior | 312 | 226 | 86 | NA | NA | 35% | 19% |
| Senior | 231 | 169 | 62 | NA | NA | 26% | 14% |

**Table 2.** Summary information for the collected covariates taken across the online (OL) and in-person (IP) groups.

| Term | N | Score Mean (SD) | $N_{OL}$ | $N_{IP}$ | OL Score Mean (SD) | IP Score Mean (SD) | IP-OL Score Mean | $ATE_{MLR}$ (95% CI) | $ATE_{IPW}$ (95% CI) | $ATE_{DR}$ (95% CI) |
|---|---|---|---|---|---|---|---|---|---|---|
| F17 | 271 | 87.27 (8.8) | 135 | 136 | 86.52 (9.13) | 88.01 (8.41) | -1.49 (1.07 | -1.74 (-3.48,0. | -1.14 (-2.73,0. | -1.55 (-3.24,0. |
| Sp1 8 | 27 4 | 85.36 (10.3 | 15 8 | 11 6 | 86.88 (9.44) | 83.29 (11.01) | 3.59 (1.26 | 1.75 (-0.42,4. | 1.82 (-0.37,4. | 1.71 (-0.53,4. |
| Sp1 9 | 26 7 | 83.30 (10.8 | 20 1 | 66 | 84.64 (9.90) | 79.19 (12.36) | 5.45 (1.53 | 2.07 (-0.78,4. | 2.17 (-0.60,5. | 2.22 (-0.55,5. |
| F19 | 29 3 | 86.65 (10.5 | 15 3 | 14 0 | 87.44 (10.26) | 85.78 (9.94) | 1.56 (1.22 | 1.19 (-0.91,3. | 1.39 (-0.83,3. | 1.05 (-1.09,3. |

**Table 3.** Raw exam scores and their SDs are given for online (OL) and in-person (IP) groups along with calculated SE of their difference and calculated ATEs using three different adjustment methods. The ATE calculations used all the covariates except ACTverbal, which was predictable with $R^2$=0.89 from the other covariates, and whose inclusion would have a negligible effect (+0.02) on the $ATE_{MLR}$.

| Students | ATE_MLR | ATE_IPW | ATE_DR | ATE_OHAL |
|---|---|---|---|---|
| All | 0.63<br>(-0.44, 1.71) | 0.75<br>(-0.34, 1.86) | 0.58<br>(-0.44, 1.63) | 0.63<br>(-0.50, 1.75) |
| With HSGPA adjustment | 0.74 ±1.16 | 0.83 ±1.15 | 0.73 ±1.10 | |

**Table 4** ATE estimates and confidence intervals using the four different adjustment methods and after adjustment for inclusion of HSGPA in the model.

| Variable | slope | 95% Confidence interval |
|---|---|---|
| **Intercept** | 29.48 | (22.46, 36.95) |
| **Online** | 0.63 | (-0.44, 1.72) |
| **Female** | 0.41 | (-0.59, 1.37) |
| **International** | 0.86 | (-0.18, 2.01) |
| **F17** | 0.05 | (-1.32, 1.38) |
| **S18** | -1.09 | (-2.47, 0.42) |
| **S19** | -2.60 | (-4.07, -1.17) |
| **FR** | -1.90 | (-3.78, 0.01) |
| **SO** | -0.27 | (-1.69, 1.14) |
| **JR** | -0.50 | (-1.83, 0.91) |
| **ACTMajor** | 0.518 | (0.298, 0.753) |
| **ACT** | 0.224 | (-0.023, 0.497) |
| **ACTMath** | 1.065 | (0.810, 1.303) |

**Table 5** The least-squares point estimates for the predictive coefficients in the multiple linear regression model based on the 1105 student sample are shown. The 95% confidence intervals are calculated by a nonparametric bootstrap method, and are close to those obtained using a t distribution. The intercept represents a hypothetical domestic male senior in the in-person F19 class with 0's for all ACT scores.